\documentclass[12pt,a4paper]{article}
\usepackage{amsfonts}
\usepackage{amssymb}
\usepackage{physics}
\usepackage{tikz}
\usepackage[left=1cm,right=1cm,top=1cm,bottom=2cm]{geometry}
\usepackage{setspace}
\usepackage{booktabs}\usetikzlibrary{shapes.callouts}
\usepackage[numbers,sort&compress]{natbib}

\usepackage{pgfplots}

\pgfplotsset{compat = newest}

\tikzset{
    level/.style = {
        line width=4pt,
        black,
    },
    connect/.style ={        dashed,
        red
    },
    notice/.style = {
        draw,
        rectangle callout,
        callout relative pointer={#1}
    },
    label/.style = {
        text width=2cm
    }
}

\author{M.~O.~Musa$^1$, H.~Temimi$^1$ and Y. Zhu$^2$\\
$^1$Department of Mathematics and Natural Sciences\\
 Gulf University for Science and Technology\\ 
  P.~O.~Box  7207 Hawally 32093, Kuwait\\
  $^2$Department of Physics\\
  Florida International University\\
  Miami, FL, U.S.A.}
\title{Semiclassical and quantum nonlinear spectra of a strongly coupled single $\Lambda$-type three-level atom-cavity QED system}
\begin{document}

\doublespacing

\tikzset{>=latex}

\maketitle

\begin{abstract}
We present detailed numerical simulations of semiclassical and quantum spectra of a cavity quantum electrodynamics system consisting of a single three-level  atom in $\Lambda$-configuration with one of its transitions strongly interacting with a quantized cavity mode while the other is driven by a coherent classical  field.  After  deriving the equations of motion for the expected values of the system operators from the master equation, we compute numerically the semiclassical and quantum spectra of the system under various levels of external driving field strengths. In the semiclassical approach we neglect the quantum correlations between cavity and atomic operators, while we make no such assumption in the fully quantum approach. We show that, under sufficiently weak driving field conditions, the semiclassical and fully quantum mechanical approaches result in identical spectra. However at higher driving field intensities, the two approaches yield starkly different results: The fully quantum mechanical approach results in  multiphoton spectrum with well-defined structure while the semiclassical results in a bistable spectrum.  Our results also reveal that the Raman transition mediated by the dark state of the system has a complex structure that depends on the manner in which the system is probed.
\end{abstract}
\section{Introduction}
Strong coupling between atomic transitions and quantized cavity modes creates an entangled, inherently nonlinear cavity quantum electrodynamics (QED) system. The simplest system used to study such interactions is the well-studied Jaynes-Cumminngs (JC) model~\cite{jaynes63}. This mathematically and conceptually simple model captures fully  spectral characteristics of interacting two-level atom and a quantized cavity mode and is one of the most-studied models in quantum optics~\cite{haroche92,shore93,haroche07,zhu88, raizen89, zhu90}. Replacing the two-level atom with  three-level atom adds complexity to the structure and optical characteristics cavity QED system. Three level atoms in free space driven by coherent classical fields has been the subject great deal studies and led to  the discovery of many interesting applications, such as electromagnetically induced transparency~\cite{harris98, marangos98, boller91, eberly94, budker99, hau99, phillips01, ginsberg07} and coherent population control atoms and molecules~\cite{kuklinski89, marte91,vitanov01} and powerful applications in the time domain such as stimulated Raman adiabatic passage (STIRAP)~\cite{unanyan98, marte91,bergmann15, vitanov17}. 
More recently, the focus has shifted on three-level systems interacting quantized cavity modes mainly in the context of all-optical switches~\cite{zhang07, yang14,zou14} and intra-cavity nonlinear optics~\cite{kwon13}. Earlier studies have been carried out in the so-called weak coupling regime where the atom-cavity coupling constant value is less than the dissipation rates of the excited atomic state and cavity mode. Under such a regime, the correlations between atomic and cavity operators are not very important and are usually neglected in what is termed as semiclassical approach. However,  recent experiments carried out in the optical domain have reached cavity quality values that place the atom-cavity coupling constant firmly in the so-called strong coupling regime  where the atom-cavity correlations are no longer negligible~\cite{brune96, schuster08,koch11, ourjoumtsev11,sames14, hamsen17, villas-boas20,tollazi21} and observation of multiphoton excitations is possible. These experiments open up new opportunities for new applications of cavity QED systems as well detail theoretical understanding of the behavior of the system.  Among other things, there is a strong need for fully understanding the energy structure and spectral characteristics of driven strongly interacting three-level atom-cavity QED systems. In this work, we aim to demonstrate the full spectrum under semiclassical and fully quantum mechanical approaches and how the semiclassical analysis is inadequate under sufficiently strong external driving field conditions. We also explore subtleties  of the Raman processes mediated by the dark dressed state of the system under sufficiently strong driving fields.
\section{Theoretical model}
We consider a single, motionless three-level atom in $\Lambda$-configuration consisting of one excited state $\ket{3}$ and two lower states $\ket{1}$ and $\ket{2}$. The $\ket{3}\leftrightarrow\ket{1}$ transition  of the atom is strongly coupled with a quantized, externally driven cavity mode, whereas  the $\ket{3}\leftrightarrow\ket{2}$ is driven by a coherent classical field. We also consider the cavity to be lossy with dissipation rate $\kappa$ and  the free-space spontaneous emission of the atomic transitions to be $\gamma_{3i}$, where $i=2,1$. We further assume that the $\ket{2}\leftrightarrow\ket{1}$ is forbidden (i.e., $\gamma_{21}=0$). The schematic diagram of the system is shown in Figure~\ref{fig:schematic}. The Hamiltonian  of the system in the frame co-rotating with the laser field is 
\begin{equation}
H=\hbar\Delta_c a^{\dagger}a+\hbar\Delta_{1}\sigma_{31}\sigma_{13}+\hbar(\Delta_1-\Delta_2)\sigma_{21}\sigma_{12}+\hbar(g\sigma_{31}a+\Omega\sigma_{32}+\varepsilon a + \text{H. c.})
\end{equation}
where H.c. stands for Hermitian conjugate, $\varepsilon$ represents the strength of driving field, $a$ ($a^{\dagger}$) is the
cavity mode annihilation (creation) operator and $\sigma_{ij}\equiv \ket{i}\bra{j}$ ($\sigma_{ji}= \sigma_{ij}^{\dagger}$) ($i=1, 2, 3$; $i\ne j$) are the atomic state inversion operators. In addition, $\omega_{31} = (E_3 - E_1)/\hbar$ is the transition frequency of the $\ket{3}\leftrightarrow\ket{1}$ transition and $\Delta_i=\omega_{3i}-\omega_l$ ($i=1,2$)  and  $\Delta_c=\omega_c-\omega_l$ are the detunings of the atomic transitions and the cavity mode from the external driving frequency $\omega_l$, respectively. The coupling parameter $g=\sqrt{\omega_c/2\hbar\epsilon_0V}\mu_{31}$, where $\mu_{31}$ is transition dipole moment and $\omega_c$ is the cavity resonance frequency, controls the strength of the coupling between the cavity mode and the atomic transition. In addition $\Omega=\mu_{32}E_0$, where $E_0$ is the electric field amplitude of the laser field and $\mu_{32}$ is transition dipole moment, represents the strength of the interaction between the second transition and the free-space laser field. The system is the weak coupling regime for values of $g/2\kappa\lambda\lesssim1$,  whereas it is in the strong coupling regime if $g/2\kappa\lambda\gg 1$.
\subsection{Dressed states of the undriven system}
The matrix representation of the Hamiltonian of the interacting system in the uncoupled basis \{$\{\ket{1;n+1},\ket{2;n}, \ket{3;n}\}$\} has the  form 
\begin{equation}
H=\hbar
\begin{pmatrix}
(n+1)\Delta_c & 0 & g\sqrt{n+1}\\
0 & n\Delta_c+\Delta_1-\Delta_2 & \Omega \\
g\sqrt{n+1} & \Omega & n\Delta_c+\Delta_1
\end{pmatrix}.
\label{eq:matrix_form}
\end{equation}
For arbitrary detunings, the eigenvalues of the Hamiltonian are complicated functions of the coupling constants and the detunings which reveal very little information about the nature of spectrum of the system. However, in the case $\Delta_2=\Delta_1$ the eigenfrequencies and the eigenvectors of the system take the forms:
\begin{eqnarray}
\omega^{0}_{n} &=& \omega_c(n+\tfrac{1}{2})\nonumber \\
\omega^{\pm}_{n}&=&\omega_c(n+\tfrac{1}{2})+\frac{1}{2}\left((\omega_{31}-\omega_c)\pm\sqrt{4g^2(n+1)+\Omega^2+(\omega_{31}-\omega_c)^2}\right) 
\label{eq:dressed_state_energy}
\end{eqnarray} 
and 
\begin{eqnarray}
\label{eq:dressed_states}
\ket{0,n}&=&-\sin\theta_n\ket{1,n+1}+\cos\theta_n\ket{2,n} \nonumber \\ 
\ket{-,n}&=&-\cos\theta_n\cos\phi_n\ket{1,n+1}-\sin\theta_n\cos\phi_n\ket{2,n}+\sin\phi_n\ket{3,n}\nonumber \\ 
\ket{+,n} &=& \cos\theta_n\sin\phi_n\ket{1, n+1}+\sin\theta_n\sin\phi_n\ket{2,n}+\cos\phi_n\ket{3,n}. 
\end{eqnarray}
The set of equations in Eq.~\ref{eq:dressed_states} is called the dressed states of the system and the so-called mixing angles $\theta_n$ and $\phi_n$ are defined in terms of the system parameters as
\begin{equation}
\tan\theta_n = \frac{\Omega}{2g\sqrt{n+1}}
\label{eq:theta}
\end{equation}
and
\begin{equation}
\tan \phi_n = \frac{\sqrt{4g^2(n+1)+\Omega^2}}{(\omega_c-\omega_{31})+\sqrt{4g^2(n+1)+\Omega^2+(\omega_{31}-\omega_c)^2}}.
\label{eq:phi}
\end{equation}
In the absence of the classical field (i.e., $\Omega=0$), $\sin\theta\rightarrow 0$ and $\cos\theta\rightarrow 1$ and the system decouples into a two-level cavity QED system  and an isolated state, namely, $\ket{0,n}\rightarrow \ket{2,n}$. 
\subsection{Open system dynamics}

The Hamiltonian of the system accounts for the structure and the time evolution of the system in the absence of dissipation.  However, analysis of the realistic behavior of the system requires taking into account the effect of dissipation on the evolution of the system. The density matrix formalism offers a natural way of taking dissipation. The behavior of the system may be extracted from the density matrix of the system  which obeys the master equation
\begin{equation}
\dot{\rho} = -\frac{i}{\hbar}\comm{H}{\rho}+\tfrac{1}{2}\kappa\mathcal{L}_c\rho+\tfrac{1}{2}\sum_{j=1}^2\gamma_{3j}\mathcal{L}_{j}\rho
\label{eq:master_equation}
\end{equation}
where  $\kappa$ and $\gamma_{3j}$ ($j=1,2$)are the cavity and excited state decay rates and 
\begin{equation}
\mathcal{L}_c\rho=2a\rho a^{\dagger}-a^{\dagger}a\rho-\rho a^{\dagger}a 
\end{equation}
represents the cavity damping whereas
\begin{equation}
\mathcal{L}_j\rho=2\sigma_{j3}\rho \sigma_{3j}-\sigma_{3j}\sigma_{j3}\rho-\rho \sigma_{3j}\sigma_{j3}
\end{equation}
represents the decay of the excited state $\ket{3}$ into the lower atomic atomic states $\ket{j}$ ($j=1,2$). In the derivation of Equation~\ref{eq:master_equation}, interactions between the cavity QED system and the environment are treated as first order (Born approximation), memoryless stochastic interactions~\cite{carmichael99}. 
\subsection{Equations of motion}
The time evolution of the atomic operators is specified by  following equations of motion:
\begin{eqnarray}
\ev{\dot{\sigma}_{11}} &=& ig\ev{\sigma_{31}a}-ig\ev{\sigma_{13}a^{\dagger}}+\gamma_{31}\ev{\sigma_{33}}-i\Omega\ev{\sigma_{32}}+i\Omega\ev{\sigma_{23}} \nonumber \\
\ev{\dot{\sigma}_{12}} &=& i(\Delta_1-\Delta_2)\ev{\sigma_{12}}+ig\ev{\sigma_{32}a} - i\Omega\ev{\sigma_{13}} \nonumber \\
\ev{\dot{\sigma}_{13}} &=& i(\Delta_i+\tfrac{1}{2}\gamma_{31})\ev{\sigma_{13}}+ig\ev{\sigma_{33}a}-ig\ev{\sigma_{11}a}-i\Omega\ev{\sigma_{13}} \nonumber \\
\ev{\dot{\sigma}_{23}} &=& i(\Delta_2+\tfrac{i}{2}(\gamma_{31}+\gamma_{32}))\ev{\sigma_{23}}+i\Omega(\ev{\sigma_{33}}-\ev{\sigma_{11}})+ig\ev{\sigma_{21}a} \nonumber \\
\ev{\dot{\sigma}_{22}} &=& \gamma_{32}\ev{\sigma_{33}} + i\Omega(\ev{\sigma_{32}}-\ev{\sigma_{32}}),
\label{eq:coupled_eqs}
\end{eqnarray}
where $\sigma_{ij}\equiv\ket{i}\bra{j}$ and  $\sigma_{11}+\sigma_{22}+\sigma_{33}=1$, whereas the cavity field evolves according to
\begin{equation}
\ev{\dot{a}}=-i(\Delta_c-\tfrac{i}{2}\kappa)\ev{a}+ig\ev{\sigma_{13}}+i\varepsilon
\label{eq:photon_eq}
\end{equation}
with the atomic population subject to the population conserving constrain: $\sigma_{11}+\sigma_{22}+\sigma_{33}=1$.
\subsection{Steady-state solutions}
In this work, we are interested in the response of the driven atom-cavity system in the frequency domain under steady state conditions. The system reaches steady state conditions after all the transients die out in time scale determined by the atomic and cavity dissipation rates. Under such conditions, all the time derivatives vanish, that is  $\ev{\dot{\sigma}_{ij}}= \ev{\dot{a}}\approx 0$ and Equations~\ref{eq:coupled_eqs} and~\ref{eq:photon_eq} reduce to a coupled set of nonlinear algebraic equations. In general, the nonlinearies in the system's response arise from correlations between the atomic and cavity operators in such terms as $\ev{\sigma_{31}a}$ and $\ev{\sigma_{11}a}$, etc. We solve this system equations under two very different assumptions. In the semiclassical case, we neglect the correlations between the cavity and atomic operators. This assumption is equivalent to replacing expectations of atomic and cavity operator products such as $\ev{\sigma_{31}a}$ with products of expectations, namely,  $\ev{\sigma_{31}}\ev{a}$, etc., and explore the consequences of the this assumption for the frequency response of the system. In the fully quantum mechanical case, we solve the system without making such assumption. In order to compute the steady state response of the system, we compute the steady density matrix which we subsequently use compute expectation values any of the atomic and cavity operators and their products using the numerical recipes outlined in Refs.~\cite{tan99, johansson12}.
\section{Results}
In what follows, we present the results of semiclassical and fully quantum mechanical simulations of the spectrum of the system. In the semiclassical case, the transition intensity of the system is proportional to the cavity field intensity, namely, $|a|^2$, and to cavity photon number $\ev{a^{\dagger}a}$ in the fully quantum case. For simplicity, we use these quantities, namely, $|a|^2$ and $<a^+a>$, in the following plots instead of the transmission intensities.

\subsection{Semiclassical case}

The semiclassical analysis of the response of the three-level atom-cavity system is based on the assumption that, regardless of the strength of coupling between the cavity and the atom or the strength of the driving field, the coherences between atomic and cavity operators are negligible. Mathematically, this assumption is equivalent to replacing expectation values of products of cavity and atomic operators with products of  their expectations. For example, the expectation $\ev{\sigma_{31}a}$, which represents the simultaneous excitation of the atom and destruction of a cavity photon, is replaced with the product these operators, namely, $\ev{\sigma_{31}}\ev{a}$. The nonlinearities in the response of the system under the semiclassical assumption are the results of the presence of these products in the equations of motion of the system. In the following, we explore the effect of this assumption on the spectrum of the system at different levels of external excitation. We choose the  atomic and cavity damping parameters in relation to the of the atom-cavity coupling strength as $\gamma_{31}/g=\gamma_{32}/g=\kappa/g=0.0083$ as we also use up to 50 cavity levels to insure accuracy of the calculations. This choice of parameters places the system deeply in the strong coupling regime. 

\subsubsection{Three-level semiclassical spectrum}
\begin{figure}
\includegraphics[width=\textwidth]{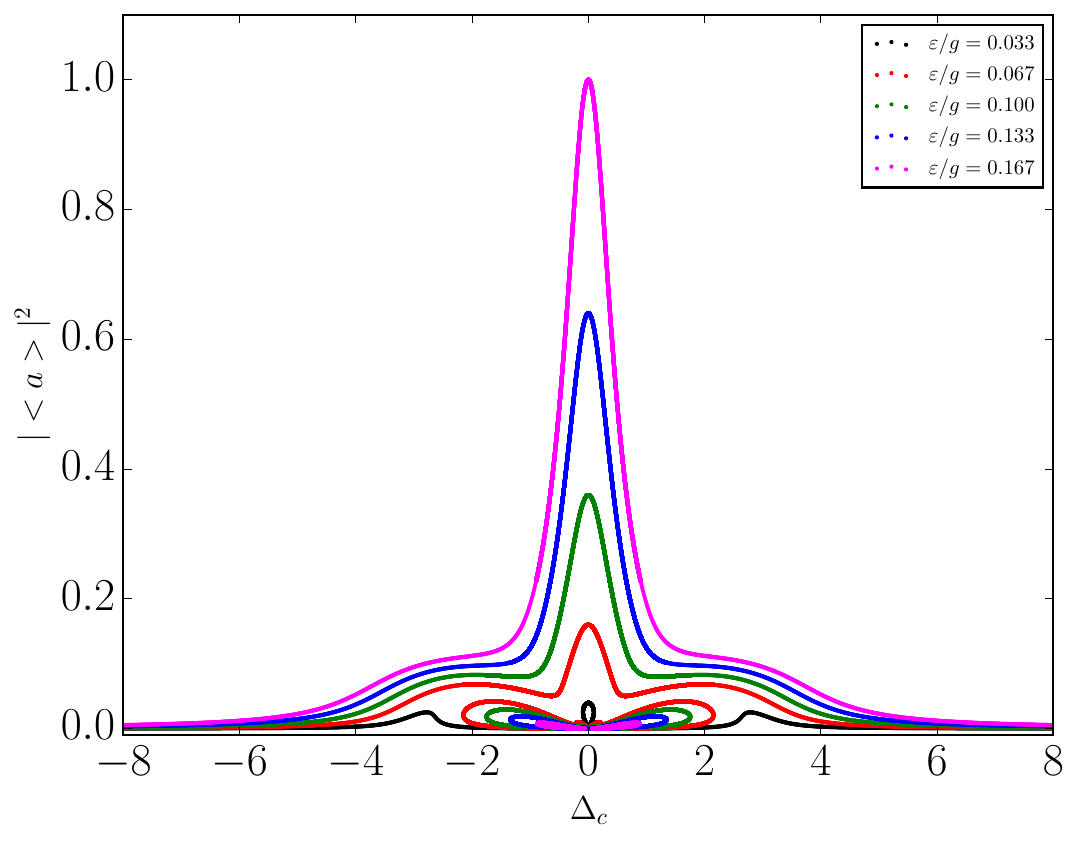}
\caption{Semiclassical spectra of the three-level cavity system at different driving field strengths. The curves represent the cavity field intensity as function of driving field detuning from the cavity resonance.}
\label{fig:semiclassical_spectrum}
\end{figure}

The simulated semiclassical spectrum of the system are obtained by numerically solving the nonlinear system of algebraic equations resulting from Equations~\ref{eq:coupled_eqs} and~\ref{eq:photon_eq} with the time derivatives set to zero (steady-state condition).  Figure~\ref{fig:semiclassical_spectrum} shows the transmission spectrum of the system under $\Omega/g=0.033$ for $\varepsilon/g$ ranging from  $0.033$, to $0.167$. The spectrum of the system is obtained by repeatedly solving the steady-state algebraic equations while $\Delta_c$ is incremented. As shown in  Figure~\ref{fig:semiclassical_spectrum}, under very low driving field conditions, the spectrum of the system consists of three peaks, which correspond to the dressed states of the system. The middle peak corresponds to coupling between the two lower states mediated by the dark state as a result of the coupling between the dark state $\ket{0,n}$ and the decoupled state $\ket{2,n}$. The peak at the center is absent when the classical field is off (i.e., $\Omega=0$) as the the dark state is not excitable through the cavity mode.  At higher driving fields, the peaks grow as their shapes start begin distorted when the classical strength $\Omega/g$ reaches $\approx 0.08$.  Further increases of the strength of the classical field  cause the distortion of the peaks to become more pronounced as they develop into connected lobes with maxima around original peak positions. As $\Omega/g$ increases further, the lobes merge  into a single strong central peak at the atom-cavity resonance. A notable feature of the semiclassical spectra of the three-level atom-cavity system is the absence of any indication of existence of multiphoton transitions.

\subsubsection{Optical bistability}
\begin{figure}
\includegraphics[width=\textwidth]{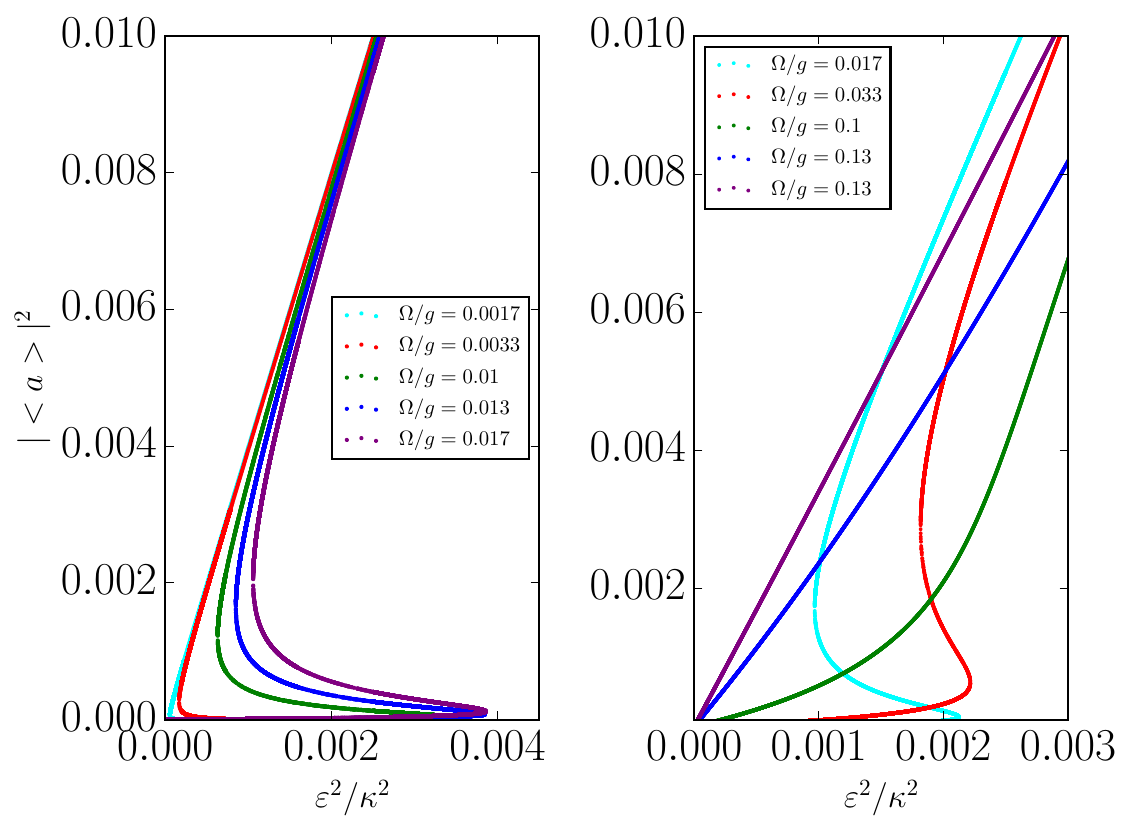}
\caption{Bistable behaviour of the three-level cavity QED system under various levels of classical field strength. The curves in the figure correspond lower values of $\Omega/g $ ranging from 0.0017 to 0.017 whereas the curves in the figure on the right result from higher values of $\Omega/g$ ranging from 0.017 to 0.17.}
\label{fig:bistability}
\end{figure}

The development of lobes in the plots in Figure~\ref{fig:semiclassical_spectrum} indicates existence of a bistable behavior similar to that of two-level cavity  QED systems~\cite{musa19}. An optical system is said to be bistable if it has two output intensities corresponding to the same input intensity over a range of input intensity values~\cite{gibbs85}.  To confirm existence of bistability in this system, we numerically solve the Equations~\ref{eq:coupled_eqs} (in the steady state limit) for the cavity mean photon number as function of the driving field intensity  for  a different values of $\Omega/g$ under resonant atom-cavity conditions. The results of these simulations are summarized in Figure~\ref{fig:bistability}. As can be inferred from the figure, the bistable behavior starts to appear around $\varepsilon/g\approx 0.017$ and becomes more pronounced as $\Omega/g$ increases. We note that the bistable behavior of the system disappears for values of $\Omega/g\gtrsim 0.067$ as the system shows cavity-like behavior and the spectrum of the system reducing to a single wide central peak at the atom-cavity resonance.
\subsection{Quantum approach}
Steady state expectation values of atomic and cavity operators, such as mean cavity photon number $\ev{a^{\dagger}a}$, atomic state populations, etc., are calculated at a specific driving frequency from the numerically-computed steady state density operator. That is, the steady state expectation of any system operator $\mathcal{\hat{O}}$ is computed from the steady state density matrix through the trace operation $\text{Tr}(\mathcal{\hat{O}}\rho_s)$ and its spectrum  is calculated by repeating this process while the driving field frequency is incrementally changed. To illustrate the detailed spectrum of the coupled atom-cavity system, we chose the following system parameters values relative to the atom-cavity coupling constant $g$: free-space field strength either $\Omega=0$ (off) or $\Omega/g\approx 0.017$, cavity and atomic dissipation rates  of $\kappa/g=\gamma_{31}/g=\gamma_{32}/g=3.3\times 10^{-4}$, atomic dephasing rate of $\gamma_{21}=0$. These parameter values correspond cooperativity parameter value of  $C=g^/2\gamma\kappa=4.5\times 10^6$, which places the system in the strong coupling regime.

\subsubsection{Excitation schemes}
The spectrum of the system depends on the manner in which the system is probed. There are three ways the system may be probed and the structure of the spectrum depends strongly on the choice of the probing scheme. As illustrated in Figure~\ref{fig:disp_curves}, the three different schemes correspond to different ways of traversing in the ($\Delta_a,\Delta_c$) space. The dotted double arrows in Figure~\ref{fig:disp_curves} represent the different probing direction under resonant atom-cavity conditions. The resonances of the system occur at the intersections of the dotted double arrows with the dressed state curves. The simplest and most practical of the three schemes is the  so-called diagonal scanning scheme in which the driving field frequency is swept while holding cavity frequency constant.  This is equivalent to traversing the ($\Delta_a,\Delta_c$) space in Figure~\ref{fig:disp_curves}  diagonally and,  as shown in Figure~\ref{fig:two_level_spec}-(a) and (b), the resulting spectra are symmetrical about the cavity resonance frequency, i.e., $\Delta_c=0$. In general, any mismatch between the atomic and cavity resonance frequencies  results in an asymmetrical spectra as illustrated in Figure~\ref{fig:two_level_spec}-(c) and (d). The larger the mismatch between the cavity and atomic resonances is, the stronger the asymmetry in the spectrum. 

The two other scanning schemes are not as simple but nevertheless represent alternative probing possibilities for the system. In the so-called vertical scanning scheme, the cavity frequency is swept while the the probe frequency is locked to the atomic resonance. This is equivalent to traversing in the ($\Delta_a,\Delta_c$) space in Figure~\ref{fig:disp_curves} vertically (dotted vertical double arrow). The resulting spectrum  under this scheme (in case $\Omega=0$) is shown in Figure~\ref{fig:horiz_vertical_scan}-(a), (b). Figure~\ref{fig:horiz_vertical_scan}-(a) corresponds to the case the external driving field is locked to the atomic resonance, whereas Figure~\ref{fig:horiz_vertical_scan}-(b) corresponds to the case where the external driving field is detuned from the atomic transition by the amount $\Delta_a/g=1$. A notable feature of the spectrum of the system under this scanning method is the absence from the spectrum by the broader single photon peaks. The reason for this is, as can be seen in Figure~\ref{fig:disp_curves} the lack of intersection between the single photon dressed state energy and that of the cavity photons. Single photon transitions are forbidden under this probing method.

In the third scheme, the probe field is locked to the cavity resonance frequencies as they are scanned together over the atomic resonance (alternatively, the atomic resonance may be tuned while holding the cavity and driving field frequencies fixed).  This scheme is equivalent to traversing  horizontally in the ($\Delta_a,\Delta_c$) space. Figure~\ref{fig:horiz_vertical_scan} (c)-(d) shows the spectrum resulting from this scanning scheme. In particular, if resonance between the driving field and the cavity mode is maintained during scanning, no excitations occur and no spectrum is produced as shown in Figure~\ref{fig:horiz_vertical_scan}-(c). The reason for this is due to lack of intersections between the atomic energy and the dressed state energy curves. Figure~\ref{fig:horiz_vertical_scan}-(d) shows one of the two branches of the system spectrum resulting when there is a sufficient mismatch between the driving field and the cavity resonance. The spacing between successive transitions is increased significantly which makes this scanning method ideal for selective multiphoton excitation of the dressed states.  

In addition to these methods, it is also possible to scan the classical field frequency of the system resonances.
\subsubsection{Two-level spectrum ($\Omega=0$)}
When the classical field off (i.e., $\Omega=0$), the third atomic state (i.e., $\ket{2,n}$) is radiatively decoupled from rest of the system. The transition amplitude between this state and any of the excited states is zero regardless of the number of photons in the cavity and there is no population transfer to it. Additionally, the excited state $\ket{0,n}$ does not radiatively couple to the ground state $\ket{1, n+1}$. This state does not contribute to the spectrum of the system and it is appropriately called the dark state for this reason. If the external driving field is tuned to the dark state transition, the system does not absorb any photons and there is no population transfer to any other state through it.  
Due to these reasons, only the dressed excited states pairs $\ket{\pm,n}$ contribute to the excitation spectrum of the system. Under this condition, the system effectively behaves as a two-level cavity QED system. The simulated multiphoton spectrum of the this effective two-level cavity QED system is shown in Figures~\ref{fig:two_level_spec}. This spectrum is similar to the multiphoton spectrum of the two-level cavity QED discussed in detail in Ref.~\cite{musa19}. 
\subsubsection{Three-level spectrum ($\Omega\ne 0$)}
We discuss above how the three-level cavity QED system becomes an effective two-level system when the classical field is off (i.e., $\Omega=0$). Next we consider the spectrum of the system under a relatively weak classical field irradiation. The classical field introduces additional degrees of freedom to the system. In the following, we consider the cases where  the frequency of the classical tuned resonantly either to the dark state(i.e., $\ket{0,0}$) or to one of the other dressed states (i.e., $\ket{\pm,0}$). Figure~\ref{fig:three_level_diagonal_scan_spec} shows the simulated spectrum of the system under the diagonal scanning scheme and resonant atom-cavity conditions with the classical field tuned to the dark state. Panels (a), (c) and (e) show the mean cavity photon number and the populations of the two atomic $\ket{3}$ and $\ket{2}$ as functions of the external driving field detuning. These spectra are obtained from the same simulation run. The spectra in Panels (a) and (c) are qualitatively similar in structure to those of the two-level cavity QED in Figure~\ref{fig:two_level_spec} (a), (c), except for extra peak structure at $\Delta_c=0$. There are also slight shifts of the peak positions of the two main branches, relative to the those of the two level system, caused by the presence of the classical field. The peak structure around $\Delta_c=0$ in Figure~\ref{fig:three_level_diagonal_scan_spec} (e) represent population transfer to the decoupled lower ground state $\ket{2,0}$. This peak structure is present  only when the classical field is on and (in this case) tuned to the transition between the dark state $\ket{0,0}$ and the uncoupled state $\ket{2,n}$. The intensity of this peak structure depends on the intensity of the classical. This structure is due to dark-state mediated Raman transitions as we further discuss below their nature.

The panels on the right side of  Figure~\ref{fig:three_level_diagonal_scan_spec} (panels (b), (d) and (f)) show the mean cavity photon number and the population of the atomic state $\ket{3}$ as functions of the cavity detuning under the condition $(\Delta_{31}-\Delta_c)/g=1$. The spectra in panels (b) and (d) are qualitatively similar to those of the two-level cavity QED system in Figure~\ref{fig:two_level_spec} (panels (b) and (d)) except for the resonances near $\Delta_c=0$. These additional resonances are induced by the coupling between the dressed state $\ket{-,n}$ and the uncoupled state $\ket{2,n}$ by the classical field. Furthermore, the resonances on Figure~\ref{fig:three_level_diagonal_scan_spec}-(e) represent population transfer to the uncoupled state $\ket{2,n}$. The peak on the left represents a Raman transition mediated by the dressed states $\ket{\pm,n}$.

Figure~\ref{fig:central_peak} detailed  structure of the resonance at $\Delta_c=0$ in Figure~\ref{fig:three_level_diagonal_scan_spec}-(a), (c) and (e). Figure~\ref{fig:central_peak}-(a)  corresponds to the case where the system is scanned horizontally, whereas  Figure~\ref{fig:central_peak}-(b) corresponds to the case where the system is scanned diagonally. The existence the transition in Figure~\ref{fig:central_peak}-(a) with fringe-like structure is at first sight surprising. However, it turns out this is the result direct multiphoton Raman-type transitions between the two lower states of the system ($\ket{3,n}\leftrightarrow\ket{2,n}$). The transitions are mediated by the dark state $\ket{n,0}$ and the fringe-like structure is the result of interference of the different excitation pathways. We note that the width and resolution of the fringe-like pattern increase with the strength of the semiclassical field and disappear as it decreases to zero.

With regard to Figure~\ref{fig:central_peak}-(b) we note that the above semiclassical analysis predicts a single peak with no observable structure at $\Delta_c=0$ when the system is scanned diagonally.  However, a shown in Figure~\ref{fig:central_peak}-(b, the full quantum mechanical analysis shows existence of a more complex structure around $\Delta_c=0$. Careful examination of Figure~\ref{fig:central_peak}-(b) shows that the structure is the result of a superimposition of two types of transitions: direct multiphoton transition similar to those in Figure~\ref{fig:central_peak}-(a) and a set discrete two-level cavity QED-like transitions. While, the fringe-like transitions are the result of direct multiphoton transitions between the two lower state mediated by the dark state, the origin of the discrete transitions is different. As can be seen in Figure~\ref{fig:raman_schematic}, under resonant atom-cavity conditions, there is an interaction between the product state $\ket{1,n}$ and the dark state $\ket{0,n}$. And it is this interaction that leads to the descrete transitions. The strength of this interaction depends on the strength of the external field (or number of photons in the cavity) and is given by 
\begin{equation}
H=\hbar
\begin{pmatrix}
0 & \varepsilon\sqrt{n+1}\\
\varepsilon\sqrt{n+1} & 0 \\
\end{pmatrix},
\end{equation}
where $\varepsilon$ is the strength of the coupling between the cavity mode and the driving external field. The eigenenergies of the interaction are $\pm \hbar\varepsilon\sqrt{n+1}$, where $n$ is the number of photons in the cavity and its  eigenvectors (under resonant conditions) are  $\ket{\phi^{\pm}}=\tfrac{1}{\sqrt{2}}\ket{0,n}+\tfrac{1}{\sqrt{2}}\ket{1,n+1}$. These discrete resonances resemble those of coupled two-level cavity QED system with line positions given by the formula $\pm\varepsilon/\sqrt{n+1}$. The positions of our simulated discrete spectrum in Figure~\ref{fig:central_peak} are specified by this formula. 

\section{Discussion}
The numerical analysis presented above compares the response  of the semiclassical and fully quantum mechanical models of a $\Lambda$-type three-level atom with one its transitions coupled to a quantized cavity mode while the other is driven by a classical coherent field. The parameters for the simulations are chosen such that the system is well in the strong coupling regime. We show that at sufficiently low external driving field intensity, the two models yield very similar results: both yield a simple spectrum consisting of three peaks corresponding to the excitation of the lowest set of dressed states of the system. However, as the driving field intensity increases, the two models start yielding very different results. In the semiclassical case, the number of peaks in the spectrum stays the same, but the shapes of the peaks start getting distorted with the outer peaks developing lobes as the middle peak gets stronger and wider. The higher multiphoton resonances predicted by Equation~\ref{eq:dressed_state_energy} are completely absent from the spectrum and further analysis of the spectrum of the system shows that stronger external driving fields cause the system response to become bistable as Figure~\ref{fig:bistability} indicates.

In contrast, in the fully quantum mechanical model, higher driving field intensities lead to additional peaks in the spectrum: the stronger the driving field gets, the higher the number of peaks in the spectrum. In the simplest configuration where the atomic transition is resonant with the cavity mode (i.e., $\omega_c=\omega_{31}$) and the classical field is tuned to the dark state resonance (i.e., $\Delta_2=0$), the central "peak" is located at $\Delta_c=0$ while the two side-bands are located at $\pm\sqrt{4g^2(n+1)+\Omega^2}/(n+1)$ for $n=0,1,2,\cdots$, as shown in Figure~\ref{fig:three_level_diagonal_scan_spec}. The increase in the number of peaks in the spectrum at higher driving field intensities is the result of direct multiphoton transitions between the ground state of system and higher energy dressed states. In addition to multiphoton transitions, higher intensity driving fields cause peaks due to transitions to the lowest energy dressed states to broaden as a result of saturation.

Although simulating the spectrum of this system is far from trivial, almost all of the characteristics of the spectrum of this system (with or without the classical field)  can be predicted on the basis of the solutions of Equation~\ref{eq:matrix_form} alone. However, our simulations show that what is usually assumed a single peak around $\Delta_c=0$ has a far more complex structure. In the case the system is scanned horizontally, it has a fringe-like structure as shown in Figure~\ref{fig:central_peak}-(a). Our simulations show that the fringe-like structure of the transition gets more intense, wider and more resolved as either the driving field or the semiclassical field is increased. In the diagonally scanned case, the structure of the transition consists of a superimposed fringe-like structure and discrete structure that is similar to that  of a two-level cavity QED system with peaks positions given by $\pm \varepsilon/\sqrt{n+1}$ with $n=0,1,2,\cdots$. As we noted above, the fringe-like pattern of the transition in the diagonally scanned case is due to multiphoton interference whereas discrete transitions are the result of the resonant interaction between the $n$-photon ground state and the dark state.

\section{Conclusions}
In this work, we used semiclassical and fully quantum mechanical approached to simulate the  spectra of a single  $\Lambda$-type three-level atom with one of its transitions strongly interacting with a quantized cavity mode  while the other is driven by a coherent classical field. We show that the two approaches yield identical results under sufficiently weak driving field conditions. In addition, we show that when the driving conditions is not weak enough, the two approaches yield very different results. In the semiclassical approach, stronger driving fields lead only to distortions of the spectra while, in the quantum case, higher driving field intensities lead multiphoton excitations of higher energy dressed states. We deduce from this result that semiclassical analysis is not appropriate in cases where the atom-cavity QED system is sufficiently in the strong coupling regime and the driving is strong enough to induce multiphoton excitations. Our findings also show what is usually assumed as a single transition at the cavity resonance (under resonance atom-cavity conditions) has a complex structure that depends in the manner the system is probed. We show that, when the system is probed horizontally, this transition has a fringe-like structure that increases in width and resolution as the driving field intensity increase while, in the case the system is probed diagonally, its structure is fringe-like pattern superimposed with a discrete two-level cavity QED-like spectral lines.
\section*{Acknowledgements}
Research reported in this paper was supported by the Kuwait Foundation for the Advancement of Science  under Grant No: PR19-14SP-04.
\bibliographystyle{unsrt}

%
\newpage

\begin{figure}
\begin{tikzpicture}
\draw [fill=blue!5,very thick, rounded corners =10] (-1,-2) rectangle (17,15);
\draw[black, very thick] (1.0,12) .. controls (2.0,12.5) and (3.0,12.5) .. (4.0,12);
\draw[black, very thick] (1.0,12)-- (1.0,12.75) -- (4.0,12.75) -- (4.0,12);
\node at (2.5,13.25) {$\omega_{l_1}$};
\node at (1,10.5) {$\omega_{l_2}$};
\draw[black, very thick] (1.0,9) .. controls (2.0,8.5) and (3.0,8.5) .. (4.0,9);
\draw[black, very thick] (1.0,9) -- (1.0,8.25) -- (4.0,8.25)-- (4.0,9);
%
\draw[gray, very thick, red] (1.25,8) .. controls (2.5,10) and (2.5,11) .. (1.25,13.5);
\draw[gray, very thick, red] (3.75,8) .. controls (2.5,10) and (2.5,11) .. (3.75,13.5);
\draw[gray, very thick, red] (2.5,6.5) -- (0.75,8);
\draw[gray, very thick, red] (2.5,6.5) -- (4.25,8);
\draw[gray, very thick, red] (0.75,8) -- (1.25,8);
\draw[gray, very thick, red] (4.25,8) -- (3.75,8);
\draw[draw=black, fill=yellow, fill opacity=0.2, line width=2pt] (0.0,13.75) -- (5.0,13.75) -- (5.0,6) -- (0.0,6) -- (0.0,13.75);
\draw[gray, very thick, red] (0.5,11.25) .. controls (2.25,10.5) and (3,10.5) .. (4.0,11.25);
\draw[gray, very thick, red] (0.5,9.75) .. controls (2.25,10.5) and (3,10.5) .. (4.0,9.75);
\draw[gray, very thick, red] (4.75,10.5) -- (4.0,11.5);
\draw[gray, very thick, red] (4.75,10.5) -- (4.0,9.5);
\draw[gray, very thick, red] (4.0,9.5) -- (4.0,9.75);
\draw[gray, very thick, red] (4.0,11.25) -- (4.0,11.5);
%
\node[circle,shading=ball,minimum width=0.2cm] (ball) at (2.5,10.5) {};
\node at (2.5,5.25) {(a)};
\draw[level] (0,0) -- (2,0) node[right] {$\ket{1}$} ;
\draw[level] (2.5,1) -- (4.5,1) node[right] {$\ket{2}$} ;
\draw[level] (0,4) -- (2,4) node[right] {$\ket{e}$} ;
\node at (1,-1) {(b)};
\draw[->,>=stealth,line width=2pt,red] (0.75,0) -- (0.75,4.0)  node[black] at (1.15,2) {$\omega_{31}$};
\draw[->,>=stealth,line width=2pt,red] (3.5,1) -- (1.05,4.0)  node[black] at (2.75,2.5) {$\omega_{32}$};
\draw[level] (6,0) -- (8,0) node[right] {$\ket{0}$} ;
\draw[level] (6,4) -- (8,4);
\node at (8.3,3.6) {$\ket{1}$} ;
\draw[level] (6,8) -- (8,8);
\node at (8.3,7.6) {$\ket{2}$} ;
\draw[red, fill=blue] (7,10) circle (1.5 pt);
\draw[red, fill=blue] (7,10.5) circle (1.5 pt);
\draw[red, fill=blue] (7,11) circle (1.5 pt);
\draw[level] (6,13) -- (8,13);
\node at (7,-1) {(c)};
\draw[->,>=stealth,line width=2pt,red] (7,0) -- (7,4)  node[black] at (7.4,6.0) {$\omega_c$};
\draw[->,>=stealth,line width=2pt,red] (7,4) -- (7,8)  node[black] at (7.4,2.0) {$\omega_c$};
\draw[level] (10,0) -- (12,0) node[right] {$\ket{1,0}$} ; 
\node at (11,-1) {(d)};
\draw[level] (10,3.5) -- (12,3.5) node[right] {$\ket{0,-}$};
\draw[very thick, black] (10,3.5) -- (8,4);
\draw[level] (10,4) -- (12,4) node[right] {$\ket{0,0}$};
\draw[very thick] (10,4) -- (8,4);
\draw[level] (10,4.5) -- (12,4.5) node[right] {$\ket{0,+}$};
\draw[very thick] (10,4.5) -- (8,4);
\draw[level] (13,1) -- (15,1) node[right] {$\ket{2,0}$} ;
\draw[->,>=stealth,line width=2pt,brown] (14,1) -- (11,4) ;
\draw[->,>=stealth,line width=2pt,red] (10.5,0) -- (10.5,3.5)  node[black] at (9.7,2) {$\omega_c-g$};
\draw[->,>=stealth,line width=2pt,blue] (11.5,0) -- (11.5,4.5)  node[black] at (12.2,2) {$\omega_c+g$};
\draw[level] (10,7.29) -- (12,7.29) node[right] {$\ket{1,-}$};
\draw[very thick] (10,7.29) -- (8,8);
\draw[level] (10,8) -- (12,8) node[right] {$\ket{1,0}$};
\draw[very thick] (10,8) -- (8,8);
\draw[level] (10,8.71) -- (12,8.71) node[right] {$\ket{1,+}$};
\draw[very thick] (10,8.71) -- (8,8);
\draw[level] (13,5) -- (15,5) node[right] {$\ket{2,1}$} ;
\draw[->,>=stealth,line width=2pt,brown] (14,5) -- (11,8) ;
\draw[->,>=stealth,line width=2pt,red] (10.5,3.5) -- (10.5,7.29);  
\draw[->,>=stealth,line width=2pt,blue] (11.5,4.5) -- (11.5,8.71);  
\draw[level] (10,12) -- (12,12) node[right] {$\ket{-,n}$};
\draw[very thick] (10,14) -- (8,13);
\draw[level] (10,13) -- (12,13) node[right] {$\ket{0,n}$};
\draw[very thick] (10,13) -- (8,13);
\draw[level] (10,14) -- (12,14) node[right] {$\ket{+,n}$};
\draw[very thick] (10,12) -- (8,13);
\draw[level] (13,9) -- (15,9) node[right] {$\ket{2,2}$} ;
\end{tikzpicture}
\caption{Schematic energy diagram of the coupled three-level atom-cavity system. (a) Cavity schematic; (b) Bare atom energy structure; (c) Bare cavity energy structure; (d) Resonantly-coupled atom-cavity system energy structure  (i.e., $\omega_c=\omega_{31}$).}
\label{fig:schematic}
\end{figure}
\begin{figure}[!h]
\begin{tikzpicture}
    \begin{axis}[   restrict y to domain=-9:9, 
	xmin = -6, xmax = 6,
    ymin = -6, ymax = 6,
    xtick distance = 1,
    ytick distance = 1,
    grid = both,
    minor tick num = 2,
    major grid style = {lightgray!25},
    minor grid style = {lightgray!12.5},
    width = \textwidth,
    xlabel = {$\Delta_a$},
    ylabel = {$\Delta_c$},
	name=plot1      
    ]
        
        \addplot[
    	smooth,
    	line width=3pt,
    	cyan,
		] table[x=Da,y=Dc,col sep=comma] {zero_neg_0.csv};
		
		\addplot[
    	smooth,
    	line width=3pt,
    	cyan,
		] table[x=Da,y=Dc,col sep=comma] {zero_pos_0.csv};		
		
		\addplot[
    	smooth,
    	line width=3pt,
    	green,
		] table[x=Da,y=Dc,col sep=comma] {one_neg_0.csv};
		
		\addplot[
    	smooth,
    	line width=3pt,
    	green,
		] table[x=Da,y=Dc,col sep=comma] {one_pos_0.csv};		
		
		\addplot[
    	smooth,
    	line width=3pt,
    	blue,
		] table[x=Da,y=Dc,col sep=comma] {two_neg_0.csv};
		
		\addplot[
    	smooth,
    	line width=3pt,
    	blue,
		] table[x=Da,y=Dc,col sep=comma] {two_pos_0.csv};
		
		\addplot[
    	smooth,
    	line width=3pt,
    	magenta,
		] table[x=Da,y=Dc,col sep=comma] {three_neg_0.csv};		
		
		\addplot[
    	smooth,
    	line width=3pt,
    	magenta,
		] table[x=Da,y=Dc,col sep=comma] {three_pos_0.csv};
		
		\addplot[
		domain=-9:9,
    	smooth,
    	line width=3pt,
    	cyan,
		] {exp(0)-1};

		\draw[<->, line width=3pt, dash pattern={on 7pt off 2pt}, red] (0,-4.5)--(0,4.5);
		\node at (0,5) {$\Delta_a=0$};
		\draw[<->, line width=3pt,  dash pattern={on 7pt off 2pt}, red] (-4.5,0)--(4.5,0); 
		\node at (-4.0,0.5) {$\Delta_c=0$};
		\draw[<->, line width=3pt,  dash pattern={on 7pt off 2pt}, red] (-4.5,-4.5)--(4.5,4.5);
		\node at (5,5) {$\Delta_a=\Delta_c$};
		
		\node at (5,-0.5){$\ket{n,0}$};
		
		\node at (2.75,4.5){$\ket{0,+}$};
		\node at (-2.75,-4.5){$\ket{0,-}$};
		
		\node at (-1.5,4.5){$\ket{1,+}$};
		\node at (1.5,-4.5){$\ket{1,-}$};
		
		\node at (-3.5,3.5){$\ket{2,+}$};
		\node at (3.5,-3.5){$\ket{2,-}$};
		
		\node at (-4.5,2.2){$\ket{3,+}$};
		\node at (4.5,-2.2){$\ket{3,-}$};
		
	\end{axis}

\end{tikzpicture}			

\caption{Dressed state energy dependence on the  three-level atom-cavity resonance frequency mismatch. As the labels indicate, the two branches of curves correspond to the two sets of dressed states. The dotted double-arrows arrows indicate the different ways the system may be probed. The diagonal double-arrow corresponds to the case where the cavity and the atomic transition are in perfect resonance as the probe field is swept, whereas the vertical correspond to the case where the the probe field is locked to the cavity resonance while the atomic transition is tuned. The horizontal case corresponds to the case the external probe is locked to the atomimc transition while the cavity is tuned. }
\label{fig:disp_curves}
\end{figure}
\begin{figure}[!h]
\begin{tikzpicture}
    \begin{axis}[    
	xmin = -5, xmax = 5,
    ymin = 0, ymax = 0.00165,
    width = 0.5\textwidth,
    xlabel = {$\Delta_c$},
    ylabel = {$\ev{a^{\dagger}a}$},
	name=plot1      
    ]
        
        \addplot[
    	smooth,
    	ultra thick,
    	blue,
		] table[col sep=comma,header=false] {two-level-0-S-int.csv};
		
    	\coordinate (insetPosition) at (rel axis cs:0.94,0.65); 
		
		\draw [thick] 
        (axis cs: -3,0.00143) -- (axis cs: -1.3242, 0.00143)
        node at (-3.5,0.00133) {$n=$};
        \draw [thick] 
        (axis cs: -3,0.00143) -- (axis cs: -3, 0.00138)
        node[pos=0.5, below] {$0$}
        node at (-2.13,0.00155) {$\ket{n,-}$}
        node at (0,0.001) {$\omega_c=\omega_a$};
        
        \draw [thick] 
        (axis cs: -2.12,0.001433) -- (axis cs: -2.12, 0.00138)
        node[pos=0.5, below] {$1$};
		
        \draw [thick] 
        (axis cs: -1.732,0.00143) -- (axis cs: -1.732, 0.00138)
        node[pos=0.5, below] {$2$};

        \draw [thick] 
        (axis cs: -1.5,0.00143) -- (axis cs: -1.5, 0.00138)
        node[pos=0.6, below] {$.$};

        \draw [thick] 
        (axis cs: -1.342,0.00143) -- (axis cs: -1.342, 0.00138)
        node[pos=0.6, below] {$..$};
        
        
        \draw [thick] 
        (axis cs: 3,0.00143) -- (axis cs: 1.3242, 0.00143);
        \draw [thick] 
        (axis cs: 3,0.00143) -- (axis cs: 3, 0.00138)
        node[pos=0.5, below] {$0$}
        node at (2.13,0.00155) {$\ket{n,+}$};
        
        \draw [thick] 
        (axis cs: 2.12,0.00143) -- (axis cs: 2.12, 0.00138)
        node[pos=0.5, below] {$1$};
		
        \draw [thick] 
        (axis cs: 1.732,0.00143) -- (axis cs: 1.732, 0.00138)
        node[pos=0.5, below] {$2$};
                
        \draw [thick] 
        (axis cs: 1.5,0.00143) -- (axis cs: 1.5, 0.00138)
        node[pos=0.45, below] {$.$};
                
        \draw [thick] 
        (axis cs: 1.342,0.00143) -- (axis cs: 1.342, 0.00138)
        node[pos=0.45, below] {$..$};
        
        \node at (4.2,0.0015) {\textcolor{blue}{(a)}};
        
    \end{axis}
    \begin{axis}[    
	xmin = -5, xmax = 5,
    ymin = 0, ymax = 0.0011,
    width = 0.5\textwidth,
    xlabel = {$\Delta_c$},
    ylabel = {$\ev{\sigma_{33}}$},
	name=plot2,
	at=(plot1.right of south east),
	anchor=left of south west      
    ]
    
        \addplot[
    	smooth,
    	ultra thick,
    	blue,
		] table[col sep=comma,header=false] {two-level-0-S-pop.csv};
		
		\draw [thick] 
        (axis cs: -3,0.00093) -- (axis cs: -1.3242, 0.00093)
        node at (-3.6,0.00085) {$n=$};
        \draw [thick] 
        (axis cs: -3,0.00093) -- (axis cs: -3, 0.00088)
        node[pos=0.5, below] {$0$}
        node at (-2.13,0.001) {$\ket{n,-}$}
        node at (0,0.0006) {$\omega_c=\omega_a$};
        
        \draw [thick, draw=red] 
        (axis cs: -2.12,0.000933) -- (axis cs: -2.12, 0.00088)
        node[pos=0.5, below] {$1$};
		
        \draw [thick]
        (axis cs: -1.732,0.00093) -- (axis cs: -1.732, 0.00088)
        node[pos=0.5, below] {$2$};

        \draw [thick] 
        (axis cs: -1.5,0.00093) -- (axis cs: -1.5, 0.00088)
        node[pos=0.6, below] {$.$};

        \draw [thick] 
        (axis cs: -1.342,0.00093) -- (axis cs: -1.342, 0.00088)
        node[pos=0.6, below] {$..$};
        
        
        \draw [thick] 
        (axis cs: 3,0.00093) -- (axis cs: 1.3242, 0.00093);
        \draw [thick] 
        (axis cs: 3,0.00093) -- (axis cs: 3, 0.00088)
        node[pos=0.5, below] {$0$}
        node at (2.13,0.001) {$\ket{n,+}$};
        
        \draw [thick] 
        (axis cs: 2.12,0.00093) -- (axis cs: 2.12, 0.00088)
        node[pos=0.5, below] {$1$};
		
        \draw [thick] 
        (axis cs: 1.732,0.00093) -- (axis cs: 1.732, 0.00088)
        node[pos=0.5, below] {$2$};

        \draw [thick] 
        (axis cs: 1.5,0.00093) -- (axis cs: 1.5, 0.00088)
        node[pos=0.45, below] {$.$};

        \draw [thick] 
        (axis cs: 1.342,0.00093) -- (axis cs: 1.342, 0.00088)
        node[pos=0.45, below] {$..$};
        
        \node at (4.2,0.001) {\textcolor{blue}{(b)}};
    \end{axis}
    \begin{axis}[    
	xmin = -5, xmax = 5,
    xlabel = {$\Delta_c$},
    ylabel = {$\ev{\sigma_{33}}$},
	name=plot4,
	at=(plot2.below south west),
	anchor=above north west     
    ]
       
        \addplot[
    	smooth,
    	very thick,
    	blue,
		] table[col sep=comma,header=false] {two-level-1-S-pop.csv};
        \node at (-1,0.0009) {$\omega_c-\omega_a=g/3a$};

        \node at (4,0.00085) {(\textcolor{blue}{d)}};
    \end{axis}
    \begin{axis}[    
	xmin = -5, xmax = 5,
    xlabel = {$\Delta_c$},
    ylabel = {$\ev{a^{\dagger}a}$}, 
	at=(plot4.left of south west),
	anchor=right of south east    
    ]
        
        \addplot[
    	smooth,
    	very thick,
    	blue,
		] table[col sep=comma,header=false] {two-level-1-S-int.csv};
        \node at (2,0.0024) {$\omega_c-\omega_a=g/3$};

        \node at (4,0.0031) {\textcolor{blue}{(c)}};
    \end{axis}    
\end{tikzpicture}    
\caption{The spectrum Coupled two-level atom-cavity system  under the diagonal scanning scheme. The spectra in Panels (a) and (b) correspond the cavity mean photon number and the population of the excited state for the case where the atom is resonant with the cavity $\omega_{c}=\omega_{31}$,  whereas the spectra in Panels (c) and (d) correspond to the same quantities for the case where the cavity and atomic resonances are frequency-mismatched by $(\omega_c-\omega_{31})/g=1/3$. As labels indicated, the peaks in each branch of the spectra correspond to multiphoton transitions between the ground state  of the system and the excited pairs of dressed states $\ket{\pm,n}$.  The broader, outermost peaks in each branch  correspond to single-photon excitations, whereas the inner peaks represent multiphoton excitations of increasing orders. In the resonant case, the positions of the peaks are given by the resonance formula $\pm g/\sqrt{n+1}$, whereas in the frequency-mismatched case the positions are given by the modified resonance forumula $\pm\frac{1}{2}\sqrt{4g^2(n+1)+(\omega_{31}-\omega_c)^2}/(n+1)$.}
\label{fig:two_level_spec}
\end{figure}
\begin{figure}[!h]
\begin{tikzpicture}
    \begin{axis}[    
    width = 0.5\textwidth,
    xlabel = {$\Delta_c$},
    ylabel = {$\ev{a^{\dagger}a}$},
	name=plot1      
    ]
        
        \addplot[
    	smooth,
    	ultra thick,
    	blue,
		] table[col sep=comma,header=false] {vertical_scan_0_shift_int.csv};
		\node at (-4.5,0.0009) {(a)};
    \end{axis}
 
    \begin{axis}[at={(plot1.north east)},  
    anchor=north east,
    restrict y to domain=-6:6,
	xmin = -6, xmax = 6,
    ymin = -6, ymax = 6,
    xtick distance = 3,
    ytick distance = 3,
    grid = both,
    minor tick num = 2,
    major grid style = {lightgray!25},
    minor grid style = {lightgray!12.5},
    width = \textwidth,
    xlabel = {$\Delta_a$},
    ylabel = {$\Delta_c$},
    tiny,
    ]
        
        \addplot[
    	smooth,
    	line width=1pt,
    	cyan,
		] table[x=Da,y=Dc,col sep=comma] {zero_neg_0.csv};
		
		\addplot[
    	smooth,
    	line width=1pt,
    	cyan,
		] table[x=Da,y=Dc,col sep=comma] {zero_pos_0.csv};		
		
		\addplot[
    	smooth,
    	line width=1pt,
    	green,
		] table[x=Da,y=Dc,col sep=comma] {one_neg_0.csv};
		
		\addplot[
    	smooth,
    	line width=1pt,
    	green,
		] table[x=Da,y=Dc,col sep=comma] {one_pos_0.csv};		
		
		\addplot[
    	smooth,
    	line width=1pt,
    	blue,
		] table[x=Da,y=Dc,col sep=comma] {two_neg_0.csv};
		
		\addplot[
    	smooth,
    	line width=1pt,
    	blue,
		] table[x=Da,y=Dc,col sep=comma] {two_pos_0.csv};
		
		\addplot[
    	smooth,
    	line width=1pt,
    	magenta,
		] table[x=Da,y=Dc,col sep=comma] {three_neg_0.csv};		
		
		\addplot[
    	smooth,
    	line width=1pt,
    	magenta,
		] table[x=Da,y=Dc,col sep=comma] {three_pos_0.csv};
		
		\draw[<->, line width=1pt, 
		dash pattern={on 7pt off 3pt}, 
		red] (0,-4.5)--(0,4.5);
	\end{axis}	 

	\begin{axis}[    
    width = 0.5\textwidth,
    xlabel = {$\Delta_c$},
    ylabel = {$\ev{a^{\dagger}a}$},
	name=plot2,
	at=(plot1.right of south east),
	anchor=left of south west       
    ]
        
        \addplot[
    	smooth,
    	ultra thick,
    	blue,
		] table[col sep=comma,header=false] {vertical_scan_3_shift_int.csv};
		\node at (-4.5,0.0018) {(b)};
    \end{axis}
    
    \begin{axis}[at={(plot2.north east)},  
    anchor=north east,
    restrict y to domain=-6:6,
	xmin = -6, xmax = 6,
    ymin = -6, ymax = 6,
    xtick distance = 3,
    ytick distance = 3,
    grid = both,
    minor tick num = 2,
    major grid style = {lightgray!25},
    minor grid style = {lightgray!12.5},
    width = \textwidth,
    xlabel = {$\Delta_a$},
    ylabel = {$\Delta_c$},
    tiny,
    ]
        
        \addplot[
    	smooth,
    	line width=1pt,
    	cyan,
		] table[x=Da,y=Dc,col sep=comma] {zero_neg_0.csv};
		
		\addplot[
    	smooth,
    	line width=1pt,
    	cyan,
		] table[x=Da,y=Dc,col sep=comma] {zero_pos_0.csv};		
		
		\addplot[
    	smooth,
    	line width=1pt,
    	green,
		] table[x=Da,y=Dc,col sep=comma] {one_neg_0.csv};
		
		\addplot[
    	smooth,
    	line width=1pt,
    	green,
		] table[x=Da,y=Dc,col sep=comma] {one_pos_0.csv};		
		
		\addplot[
    	smooth,
    	line width=1pt,
    	blue,
		] table[x=Da,y=Dc,col sep=comma] {two_neg_0.csv};
		
		\addplot[
    	smooth,
    	line width=1pt,
    	blue,
		] table[x=Da,y=Dc,col sep=comma] {two_pos_0.csv};
		
		\addplot[
    	smooth,
    	line width=1pt,
    	magenta,
		] table[x=Da,y=Dc,col sep=comma] {three_neg_0.csv};		
		
		\addplot[
    	smooth,
    	line width=1pt,
    	magenta,
		] table[x=Da,y=Dc,col sep=comma] {three_pos_0.csv};
		
		\draw[<->, line width=1pt, 
		dash pattern={on 7pt off 3pt}, 
		red] (-3,-5.5)--(-3,2.5);
	\end{axis}
    \begin{axis}[    
    width = 0.5\textwidth,
    xlabel = {$\Delta_c$},
    ylabel = {$\ev{a^{\dagger}a}$},
	name=plot4,
	at=(plot2.below south west),
	anchor=above north west        
    ]
        
        \addplot[
    	smooth,
    	ultra thick,
    	blue,
		] table[col sep=comma,header=false] {horizontal_scan_3_shift_int.csv};
		\node at (-4.5,0.0009) {(d)};
    \end{axis}
    
    \begin{axis}[at={(plot4.north east)},  
    anchor=north east,
    restrict y to domain=-6:6,
	xmin = -6, xmax = 6,
    ymin = -6, ymax = 6,
    xtick distance = 3,
    ytick distance = 3,
    grid = both,
    minor tick num = 2,
    major grid style = {lightgray!25},
    minor grid style = {lightgray!12.5},
    width = \textwidth,
    xlabel = {$\Delta_a$},
    ylabel = {$\Delta_c$},
    tiny,
    ]
        
        \addplot[
    	smooth,
    	line width=1pt,
    	cyan,
		] table[x=Da,y=Dc,col sep=comma] {zero_neg_0.csv};
		
		\addplot[
    	smooth,
    	line width=1pt,
    	cyan,
		] table[x=Da,y=Dc,col sep=comma] {zero_pos_0.csv};		
		
		\addplot[
    	smooth,
    	line width=1pt,
    	green,
		] table[x=Da,y=Dc,col sep=comma] {one_neg_0.csv};
		
		\addplot[
    	smooth,
    	line width=1pt,
    	green,
		] table[x=Da,y=Dc,col sep=comma] {one_pos_0.csv};		
		
		\addplot[
    	smooth,
    	line width=1pt,
    	blue,
		] table[x=Da,y=Dc,col sep=comma] {two_neg_0.csv};
		
		\addplot[
    	smooth,
    	line width=1pt,
    	blue,
		] table[x=Da,y=Dc,col sep=comma] {two_pos_0.csv};
		
		\addplot[
    	smooth,
    	line width=1pt,
    	magenta,
		] table[x=Da,y=Dc,col sep=comma] {three_neg_0.csv};		
		
		\addplot[
    	smooth,
    	line width=1pt,
    	magenta,
		] table[x=Da,y=Dc,col sep=comma] {three_pos_0.csv};
		
		\draw[<->, line width=1pt, 
		dash pattern={on 7pt off 3pt}, 
		red] (-2.0,3.0)--(5.5,3.0);
	\end{axis}

	\begin{axis}[    
    ymin = -0.001, ymax = 0.001,
    width = 0.5\textwidth,
    xlabel = {$\Delta_c$},
    ylabel = {$\ev{a^{\dagger}a}$},
	name=plot3,
	at=(plot4.left of south west),
	anchor=right of south east      
    ]
        
        \addplot[
    	smooth,
    	ultra thick,
    	blue,
		] table[col sep=comma,header=false] {horizontal_scan_0_shift_int.csv};
		\node at (-4.5,0.0007) {(c)};
    \end{axis}
    
    \begin{axis}[at={(plot3.north east)},  
    anchor=north east,
    restrict y to domain=-6:6,
	xmin = -6, xmax = 6,
    ymin = -6, ymax = 6,
    xtick distance = 3,
    ytick distance = 3,
    grid = both,
    minor tick num = 2,
    major grid style = {lightgray!25},
    minor grid style = {lightgray!12.5},
    width = \textwidth,
    xlabel = {$\Delta_a$},
    ylabel = {$\Delta_c$},
    tiny,
    ]
        
        \addplot[
    	smooth,
    	line width=1pt,
    	cyan,
		] table[x=Da,y=Dc,col sep=comma] {zero_neg_0.csv};
		
		\addplot[
    	smooth,
    	line width=1pt,
    	cyan,
		] table[x=Da,y=Dc,col sep=comma] {zero_pos_0.csv};		
		
		\addplot[
    	smooth,
    	line width=1pt,
    	green,
		] table[x=Da,y=Dc,col sep=comma] {one_neg_0.csv};
		
		\addplot[
    	smooth,
    	line width=1pt,
    	green,
		] table[x=Da,y=Dc,col sep=comma] {one_pos_0.csv};		
		
		\addplot[
    	smooth,
    	line width=1pt,
    	blue,
		] table[x=Da,y=Dc,col sep=comma] {two_neg_0.csv};
		
		\addplot[
    	smooth,
    	line width=1pt,
    	blue,
		] table[x=Da,y=Dc,col sep=comma] {two_pos_0.csv};
		
		\addplot[
    	smooth,
    	line width=1pt,
    	magenta,
		] table[x=Da,y=Dc,col sep=comma] {three_neg_0.csv};		
		
		\addplot[
    	smooth,
    	line width=1pt,
    	magenta,
		] table[x=Da,y=Dc,col sep=comma] {three_pos_0.csv};
		
		\draw[<->, line width=1pt, 
		dash pattern={on 7pt off 3pt}, 
		red] (-4.5,0.0)--(4.5,0.0);
	\end{axis}
    
\end{tikzpicture}    
\caption{(Panels (a) and (b) show vertically scanned spectra of the system with $\Omega=0$ (effective two-level atom-cavity system) with the cavity either resonant with the atomic transition (Panel (a)) or frequency-mismatched by  $\omega_{31}-\omega_c=g/3$ (Panel (b)). The most notable feature of the spectrum under this scanning method is the absence of the single photon transitions at $\pm g$. The spectra in the lower panels correspond to the case where the system is scanned horizontally along $\Delta_c=0$ (Panel (c)) or  along  $\Delta_c=g$. Most notable feature of the spectrum under horizontal scan is the absence of any resonances in the case where $\Delta_c=0$. The reason for the absence of resonance is due to the fact that there is not intersection between the dressed state energies and $\Delta_c$.}
\label{fig:horiz_vertical_scan}
\end{figure}
\begin{figure}[!h]
\begin{tikzpicture}
    \begin{axis}[    
	xmin = -5, xmax = 5,
    xlabel = {$\Delta_c$},
    ylabel = {$\ev{a^{\dagger}a}$},
	name=plot1      
    ]       
        \addplot[
    	smooth,
    	ultra thick,
    	blue,
		] table[col sep=comma] {three-level-10-S-int.csv};
		
		\node at (-4.2,0.0025) {(a)};
    \end{axis}
    
    \begin{axis}[at={(plot1.north east)},  
    anchor=north east,
    restrict y to domain=-6:6,
	xmin = -6, xmax = 6,
    ymin = -6, ymax = 6,
    xtick distance = 3,
    ytick distance = 3,
    grid = both,
    minor tick num = 2,
    major grid style = {lightgray!25},
    minor grid style = {lightgray!12.5},
    width = \textwidth,
    xlabel = {$\Delta_a$},
    ylabel = {$\Delta_c$},
    tiny,
    ]
        
        \addplot[
    	smooth,
    	line width=1pt,
    	cyan,
		] table[x=Da,y=Dc,col sep=comma] {zero_neg_0.csv};
		
		\addplot[
    	smooth,
    	line width=1pt,
    	cyan,
		] table[x=Da,y=Dc,col sep=comma] {zero_pos_0.csv};		
		
		\addplot[
    	smooth,
    	line width=1pt,
    	green,
		] table[x=Da,y=Dc,col sep=comma] {one_neg_0.csv};
		
		\addplot[
    	smooth,
    	line width=1pt,
    	green,
		] table[x=Da,y=Dc,col sep=comma] {one_pos_0.csv};		
		
		\addplot[
    	smooth,
    	line width=1pt,
    	blue,
		] table[x=Da,y=Dc,col sep=comma] {two_neg_0.csv};
		
		\addplot[
    	smooth,
    	line width=1pt,
    	blue,
		] table[x=Da,y=Dc,col sep=comma] {two_pos_0.csv};
		
		\addplot[
    	smooth,
    	line width=1pt,
    	magenta,
		] table[x=Da,y=Dc,col sep=comma] {three_neg_0.csv};		
		
		\addplot[
    	smooth,
    	line width=1pt,
    	magenta,
		] table[x=Da,y=Dc,col sep=comma] {three_pos_0.csv};
		
		\addplot[
		domain=-6:6,
    	smooth,
    	line width=1pt,
    	cyan,
		] {exp(0)-1};
		
		\node at (-4.0,0.5) {$\Delta_c=0$};
		\draw[<->, line width=1pt,  dash pattern={on 7pt off 2pt}, red] (-4.5,-4.5)--(4.5,4.5);
				
	\end{axis}
    \begin{axis}[    
	xmin = -5, xmax = 5,
    width = 0.5\textwidth,
    xlabel = {$\Delta_c$},
    ylabel = {$\ev{a^{\dagger}a}$},
	name=plot2,
	at=(plot1.right of south east),
	anchor=left of south west      
    ]
    
        \addplot[
    	smooth,
    	ultra thick,
    	blue,
		] table[col sep=comma,header=false] {three-level-12-S-int.csv};
		
		\node at (-4.2,0.002) {(b)};
		
    \end{axis}
    
    \begin{axis}[at={(plot2.north east)},  
    anchor=north east,
    restrict y to domain=-6:6,
	xmin = -6, xmax = 6,
    ymin = -6, ymax = 6,
    xtick distance = 3,
    ytick distance = 3,
    grid = both,
    minor tick num = 2,
    major grid style = {lightgray!25},
    minor grid style = {lightgray!12.5},
    width = \textwidth,
    xlabel = {$\Delta_a$},
    ylabel = {$\Delta_c$},
    tiny,
    ]
        
        \addplot[
    	smooth,
    	line width=1pt,
    	cyan,
		] table[x=Da,y=Dc,col sep=comma] {zero_neg_0.csv};
		
		\addplot[
    	smooth,
    	line width=1pt,
    	cyan,
		] table[x=Da,y=Dc,col sep=comma] {zero_pos_0.csv};		
		
		\addplot[
    	smooth,
    	line width=1pt,
    	green,
		] table[x=Da,y=Dc,col sep=comma] {one_neg_0.csv};
		
		\addplot[
    	smooth,
    	line width=1pt,
    	green,
		] table[x=Da,y=Dc,col sep=comma] {one_pos_0.csv};		
		
		\addplot[
    	smooth,
    	line width=1pt,
    	blue,
		] table[x=Da,y=Dc,col sep=comma] {two_neg_0.csv};
		
		\addplot[
    	smooth,
    	line width=1pt,
    	blue,
		] table[x=Da,y=Dc,col sep=comma] {two_pos_0.csv};
		
		\addplot[
    	smooth,
    	line width=1pt,
    	magenta,
		] table[x=Da,y=Dc,col sep=comma] {three_neg_0.csv};		
		
		\addplot[
    	smooth,
    	line width=1pt,
    	magenta,
		] table[x=Da,y=Dc,col sep=comma] {three_pos_0.csv};
		
		\addplot[
		domain=-6:6,
    	smooth,
    	line width=1pt,
    	cyan,
		] {exp(0)-1};
		
		\node at (-4.0,0.5) {$\Delta_c=0$};
		\draw[<->, line width=1pt,  dash pattern={on 7pt off 2pt}, red] (-4.5,-3.5)--(4.5,5.5);
		
		\end{axis}
		
    \begin{axis}[    
	xmin = -5, xmax = 5,
    xlabel = {$\Delta_c$},
    ylabel = {$\ev{\sigma_{33}}$},
	name=plot4,
	at=(plot2.below south west),
	anchor=above north west     
    ]
       
        \addplot[
    	smooth,
    	ultra thick,
    	blue,
		] table[col sep=comma,header=false] {three-level-12-S-pop.csv};
		
		\node at (-4.2,0.00072) {(d)};
    \end{axis}
    \begin{axis}[    
	xmin = -5, xmax = 5,
    xlabel = {$\Delta_c$},
    ylabel = {$\ev{\sigma_{33}}$}, 
    name = plot3,
	at=(plot4.left of south west),
	anchor=right of south east    
    ]
        
        \addplot[
    	smooth,
    	ultra thick,
    	blue,
		] table[col sep=comma,header=false] {three-level-10-S-pop.csv};
        \node at (2,0.0027) {$\omega_c-\omega_a=g/3$};
		
		\node at (-4.2,0.0007) {(c)};

    \end{axis}   
    \begin{axis}[    
	xmin = -5, xmax = 5,
    xlabel = {$\Delta_c$},
    ylabel = {$\ev{\sigma_{22}}$},
	name=plot6,
	at=(plot4.below south west),
	anchor=above north west     
    ]
       
        \addplot[
    	smooth,
    	ultra thick,
    	blue,
		] table[col sep=comma,header=false] {three-level-1-s22-pop.csv};
        \node at (-1,0.0009) {$\omega_c-\omega_a=g/3$};
		
		\node at (-4.2,0.00055) {(f)};
    \end{axis}
    \begin{axis}[    
	xmin = -5, xmax = 5,
    xlabel = {$\Delta_c$},
    ylabel = {$\ev{\sigma_{22}}$}, 
    name = plot5,
	at=(plot6.left of south west),
	anchor=right of south east    
    ]
        
        \addplot[
    	smooth,
    	ultra thick,
    	blue,
		] table[col sep=comma,header=false] {three-level-0-s22-pop.csv};
        \node at (2,0.0027) {$\omega_c-\omega_a=g/3$};
		
		\node at (-4.2,0.000375) {(e)};

    \end{axis}       
    
\end{tikzpicture}  
\caption{The spectra of the three-level atom-cavity system under diagonal scanning scheme. The spectrum on Panels (a), (c) and (d) correspond to the cavity mean photon number and the populations of the atomic states $\ket{2}$ and $\ket{3}$, respectively, whereas the spectra on Panels (b), (d) and (f) correspond to the same quantities but with the cavity and atomic transitions frequency-mismatched by $\Delta_c-\Delta_{31}=g/3$. Unlike the two level case, the atomic state $\ket{2}$ is in this case populated. }
\label{fig:three_level_diagonal_scan_spec}
\end{figure}
\begin{figure}
\begin{tikzpicture}
\draw [fill=blue!5,very thick, rounded corners=10] (-1,-1) rectangle (11,10);
\draw[level] (4,0) -- (6,0) node[right] {$\ket{g,1}$} ;
\draw[level] (6.5,1.25) -- (8.5,1.25) node[right] {$\ket{2,1}$} ;
\draw[level, dotted] (4,4) -- (6,4) node[right] {$\ket{0,0}$} ;
\draw[level] (4,4.5) -- (6,4.5) node[right] {$\ket{0,+}$} ;
\draw[level] (4,3.5) -- (6,3.5) node[right] {$\ket{0,-}$} ;
\draw[level] (6.5,5.25) -- (8.5,5.25) node[right] {$\ket{2,1}$} ;
\draw[->,>=stealth,line width=2pt,blue]  (5,0) -- (5,4);
\node at (5.5,2) {$\omega_l$};
\draw[->,>=stealth,line width=2pt,red]  (7.5,1.25) -- (5,4);
\draw[level] (7,4.25) -- (9,4.25) node[right] {$\ket{\phi_0^+,0}$};
\draw[level] (7,3.75) -- (9,3.75) node[right] {$\ket{\phi_0^-,0}$};
\draw[very thick] (6,4) -- (7,4.25);
\draw[very thick] (6,4) -- (7,3.75);
\draw[level, dotted] (4,8) -- (6,8) node[right] {$\ket{0,1}$} ;
\draw[level] (4,8.75) -- (6,8.75) node[right] {$\ket{0,+}$} ;
\draw[level] (4,7.25) -- (6,7.25) node[right] {$\ket{0,-}$} ;
\draw[->,>=stealth,line width=2pt,blue]  (5,4) -- (5,8);
\node at (5.5,6) {$\omega_l$};
\draw[->,>=stealth,line width=2pt,red]  (7.5,5.25) -- (5,8);
%
\draw[level] (7,8.5) -- (9,8.5) node[right] {$\ket{\phi_1^+,1}$};
\draw[level] (7,7.5) -- (9,7.5) node[right] {$\ket{\phi_1^-,1}$};
\draw[very thick] (6,8) -- (7,8.5);
\draw[very thick] (6,8) -- (7,7.5);
\draw[level] (0,8) -- (2,8) node[right] {$\ket{1,3}$} ;
\draw[level] (0,4) -- (2,4) node[right] {$\ket{1,2}$} ;
\draw[level] (0,0) -- (2,0) node[right] {$\ket{1,1}$} ;
\draw[->,>=stealth,line width=2pt,blue]  (1,0) -- (1,4);
\node at (1.5,2) {$\omega_l$};
\draw[->,>=stealth,line width=2pt,blue]  (1,4) -- (1,8);
\node at (1.5,6) {$\omega_l$};

\node at (1,-0.75) {(a)};
\node at (5,-0.75) {(b)};
\end{tikzpicture}
\caption{Raman coupling between the dark state $\ket{0,n}$ and the uncoupled state $\ket{1,n+1}$ in the presence of the classical free-space field. (a) Uncoupled product state $\ket{1, n}=\ket{1}\otimes\ket{n+1}$ of the system (the atom is in the ground state and the cavity has $n$ photons). (b) The structure of dressed states resulting from the interaction between the dark state $\ket{0,n}$ and the product states $\ket{1,n+1}$.}
\label{fig:raman_schematic}
\end{figure}

\begin{figure}
\begin{tikzpicture}
    \begin{axis}[    
	xmin = -1, xmax = 1,
    ymin = 0, ymax = 0.01,
    xlabel = {$\Delta_c$},
    ylabel = {$\ev{a^{\dagger}a}$},
	name=plot1      
    ]
        
        \addplot[
    	smooth,
    	ultra thick,
    	blue,
    	mark repeat=5,
		] table[col sep=comma,header=false] {central_peak_horiz_int.csv};

		\node at (-0.3, 0.00055) {(a)};
    \end{axis}
    
    \begin{axis}[at={(plot1.north east)},  
    anchor=north east,
    restrict y to domain=-6:6,
	xmin = -6, xmax = 6,
    ymin = -6, ymax = 6,
    xtick distance = 3,
    ytick distance = 3,
    grid = both,
    minor tick num = 2,
    major grid style = {lightgray!25},
    minor grid style = {lightgray!12.5},
    width = \textwidth,
    xlabel = {$\Delta_a$},
    ylabel = {$\Delta_c$},
    tiny,
    ]
		\addplot[
    	smooth,
    	line width=1pt,
    	cyan,
		] table[x=Da,y=Dc,col sep=comma] {zero_neg_0.csv};

		\addplot[
    	smooth,
    	line width=1pt,
    	cyan,
		] table[x=Da,y=Dc,col sep=comma] {zero_pos_0.csv};		
		
		\addplot[
    	smooth,
    	line width=1pt,
    	green,
		] table[x=Da,y=Dc,col sep=comma] {one_neg_0.csv};
		
		\addplot[
    	smooth,
    	line width=1pt,
    	green,
		] table[x=Da,y=Dc,col sep=comma] {one_pos_0.csv};		
		
		\addplot[
    	smooth,
    	line width=1pt,
    	blue,
		] table[x=Da,y=Dc,col sep=comma] {two_neg_0.csv};
		
		\addplot[
    	smooth,
    	line width=1pt,
    	blue,
		] table[x=Da,y=Dc,col sep=comma] {two_pos_0.csv};
		
		\addplot[
    	smooth,
    	line width=1pt,
    	magenta,
		] table[x=Da,y=Dc,col sep=comma] {three_neg_0.csv};		
		
		\addplot[
    	smooth,
    	line width=1pt,
    	magenta,
		] table[x=Da,y=Dc,col sep=comma] {three_pos_0.csv};
		
		\addplot[
		domain=-6:6,
    	smooth,
    	line width=1pt,
    	cyan,
		] {exp(0)-1};
		
		\node at (-4.0,0.5) {$\Delta_c=0$};
		\draw[<->, line width=1pt,  dash pattern={on 7pt off 2pt}, red] (-4.5,0)--(4.5,0);

		\end{axis}

	\begin{axis}[    
	xmin = -0.35, xmax = 0.35,
    ymin = 0, ymax = 0.0006,
    xlabel = {$\Delta_c$},
    ylabel = {$\ev{a^{\dagger}a}$},
	name=plot2,
	at=(plot1.right of south east),
	anchor=left of south west       
    ]
        
        \addplot[
    	smooth,
    	ultra thick,
    	blue,
    	mark repeat=5,
		] table[col sep=comma,header=false] {central_peak__int.csv};

    \end{axis}
    
    \begin{axis}[at={(plot2.north east)},  
    anchor=north east,
    restrict y to domain=-6:6,
	xmin = -6, xmax = 6,
    ymin = -6, ymax = 6,
    xtick distance = 3,
    ytick distance = 3,
    grid = both,
    minor tick num = 2,
    major grid style = {lightgray!25},
    minor grid style = {lightgray!12.5},
    width = \textwidth,
    xlabel = {$\Delta_a$},
    ylabel = {$\Delta_c$},
    tiny,
    ]
      \addplot[
    	smooth,
    	line width=1pt,
    	cyan,
		] table[x=Da,y=Dc,col sep=comma] {zero_neg_0.csv};
		
		\addplot[
    	smooth,
    	line width=1pt,
    	cyan,
		] table[x=Da,y=Dc,col sep=comma] {zero_pos_0.csv};		
		
		\addplot[
    	smooth,
    	line width=1pt,
    	green,
		] table[x=Da,y=Dc,col sep=comma] {one_neg_0.csv};
		
		\addplot[
    	smooth,
    	line width=1pt,
    	green,
		] table[x=Da,y=Dc,col sep=comma] {one_pos_0.csv};		
		
		\addplot[
    	smooth,
    	line width=1pt,
    	blue,
		] table[x=Da,y=Dc,col sep=comma] {two_neg_0.csv};
		
		\addplot[
    	smooth,
    	line width=1pt,
    	blue,
		] table[x=Da,y=Dc,col sep=comma] {two_pos_0.csv};
		
		\addplot[
    	smooth,
    	line width=1pt,
    	magenta,
		] table[x=Da,y=Dc,col sep=comma] {three_neg_0.csv};		
		
		\addplot[
    	smooth,
    	line width=1pt,
    	magenta,
		] table[x=Da,y=Dc,col sep=comma] {three_pos_0.csv};
		
		\addplot[
		domain=-6:6,
    	smooth,
    	line width=1pt,
    	cyan,
		] {exp(0)-1};
		
		\node at (-4.0,0.5) {$\Delta_c=0$};
		\draw[<->, line width=1pt,  dash pattern={on 7pt off 2pt}, red] (-4.5,-4.5)--(4.5,4.5);  
        
		\end{axis}
\end{tikzpicture}    
\caption{ Raman spectrum of resonantly-coupled atom-cavity system with the free-space field on ($\Omega\ne 0$). (a) Raman spectrum when the system is scanned horizontally, (b) Raman spectrum when the system is diagonally scanned.}
\label{fig:central_peak}
\end{figure}
%

\end{document}